\documentstyle[12pt,epsfig]{article}
\newlength{\dinwidth}
\newlength{\dinmargin}
\newcommand{\ba}{\begin{array}}
\newcommand{\ea}{\end{array}}
\newcommand{\beq}{\begin{equation}}
\newcommand{\eeq}{\end{equation}}
\newcommand{\bea}{\begin{eqnarray}}
\newcommand{\eea}{\end{eqnarray}}

\def\bce{\begin{center}}
\def\ece{\end{center}}

\def\nonu{\nonumber}

\def\pa{\partial}
\def\al{\alpha}
\def\be{\beta}

\def\de{\delta}
\def\De{\Delta}
\def\ep{\epsilon}

\def\la{\lambda}
\def\La{\Lambda}

\def\om{\omega}

\def\eps6{{\displaystyle \mathop{\epsilon}^{6}}{}}

\def\nab6{{\displaystyle \mathop{\nabla}^{6}}{}}

\begin{document}
\thispagestyle{empty}
\addtocounter{page}{-1}
\begin{flushright}
KIAS-P03006 \\
{\tt hep-th/0301203}\\
\end{flushright}
\vspace*{1.3cm} \centerline{\Large \bf ${\cal N}=2$ Supersymmetric
$SO(N)/Sp(N)$ Gauge Theories  } \vskip0.3cm \centerline{\Large \bf
from Matrix Model } \vspace*{1.5cm} \centerline{{\bf Changhyun
Ahn}$^1$ and {\bf Soonkeon Nam}$^2$} \vspace*{1.0cm}
\centerline{\it $^1$Department of Physics, Kyungpook National
University, Taegu 702-701, Korea} \vspace*{0.2cm} \centerline{\it
$^2$Department of Physics and Research Institute for Basic
Sciences,} \centerline{ \it Kyung Hee University, Seoul 130-701,
Korea} \vspace*{0.8cm} \centerline{\tt ahn@knu.ac.kr, \qquad
nam@khu.ac.kr} \vskip2cm \centerline{\bf Abstract} \vspace*{0.5cm}

We use the matrix model to describe the ${\cal N}=2$
$SO(N)/Sp(N)$ supersymmetric gauge theories with massive hypermultiplets
in the fundamental representation.
By taking the tree level superpotential perturbation made of a polynomial
of a scalar chiral multiplet, the effective action for the eigenvalues of
chiral multiplet can be obtained. By varying this action with respect to
an eigenvalue, a loop equation is obtained.
By analyzing this equation, we derive the Seiberg-Witten curve
within the context of matrix model.
\vspace*{\fill}


\baselineskip=18pt
\newpage
\renewcommand{\theequation}{\arabic{section}\mbox{.}\arabic{equation}}

\section{Introduction}
\setcounter{equation}{0}

Recently Dijkgraaf and Vafa \cite{dv} have made a conjecture, the
exact superpotential and gauge couplings for a class of ${\cal
N}=1$ gauge theories can be obtained by calculating perturbative
computations in a matrix model in which the superpotential of the
gauge theory is interpreted as an ordinary potential. The earlier
works \cite{cv,dv3,dv2} motivated this conjecture. Based on this
observation, there are many works on this direction
\cite{av}-\cite{feng3}. In particular, we restrict to the
supersymetric $SO(N)/Sp(N)$ gauge theories. The model with quartic
tree level superpotential for adjoint chiral field was found
\cite{fo} and the effective superpotential was computed in the
context of matrix model and string theory on Calabi-Yau geometry
with flux. The perturbative calculation for glueball
superpotential was studied in \cite{ino}. For arbitrary tree level
superpotentials, the planar and leading nonplanar contributions
were derived by using higher genus loop equations and
diagrammatics \cite{ashoketal}. A field theoretic derivation of
the superpotential was given in \cite{jo} based on the
factorization property of Seiberg-Witten curve. An equivalence of
${\cal N}=1$ gauge theories deformed from ${\cal N}=2$ by the
addition of superpotential terms was studied with flavors
\cite{ookouchi} and without flavors \cite{feng1} based on the
Cachazo-Vafa's idea \cite{cv}: The low energy information is given
by extremization of the effective superpotential and in the field
theory analysis it is given by characterizing to the factorization
locus of Seiberg-Witten curve and the equivalence of two
description was given in \cite{cv}.

In this paper, we compute the matrix path integral over tree level
superpotential obtained from ${\cal N}=2$ SQCD by taking arbitrary
polynomial of a scalar chiral multiplet as a perturbation. The
effective theory action can be expressed as a function of an
eigenvalue of chiral multiplet. The saddle point equation implies
an algebraic equation defined on  a hyperelliptic Riemann surface.
The presence of this curve allows us to study the relation of
matrix model and the gauge theory result of perturbative
calculation. By using the basic idea of matrix model, we calculate
a partition function in terms of a glueball field, a distribution
of eigenvalue of chiral multiplet, a perturbed superpotential and
the mass of quarks. By reading off the two free energy
contributions from a partition function, one obtains the final
effective superpotential in terms of homology basis. By varying
this effective superpotential with respect to the coefficient
function appearing in the algebraic curve, one realizes the
existence of a meromorphic function on Riemann surface with the
appropriate structure of zeros and poles. In doing this, the
correct counting of the number of physical D5-branes in the
presence of orientifold planes (O5-planes) is very important
because these values determine the structure of zeros and poles
precisely. By identifying this function with the resolvent of
matrix model, the Seiberg-Witten curves for ${\cal N}=2$
$SO(N)/Sp(N)$ gauge theory with $N_f$ hypermultiplets are
rederived. For $U(N)$ gauge theory with $N_f$ flavors of quarks in
the fundamental representation, the derivation of Seiberg-Witten
curve was found in \cite{nsw2}. It would be interesting to study
the results in \cite{cdsw,seiberg,csw} to deal with the model
given this paper.

\section{$SO(N)$ matrix model }
\setcounter{equation}{0}

We will derive the Seiberg-Witten curve for ${\cal N}=2$
$SO(N)$ gauge theory with $N_f$ fundamental hypermultiplets
by computing the matrix path integral using the saddle
point method \cite{fgz,nsw2}.
Let us
consider  an ${\cal N}=2$ supersymmetric $SO(N)$ gauge theory
with $N_f$ flavors of quarks $Q^i_a(i=1, 2, \cdots, 2N_f,
a=1, 2, \cdots,
N)$ in the vector (fundamental)
representation \cite{ty,kty,aps,as,hanany,hms,dkp,aot}.
In terms of ${\cal N}=1$ superfields, ${\cal N}=2$ vector multiplet
consists of  a field strength chiral multiplet $W_{\al}^{ab}$ and a scalar
chiral multiplet $\Phi_{ab}$ both in the adjoint representation of
the gauge group.
The ${\cal N}=2$ superpotential takes the form
\bea
W_{tree}(\Phi, Q) =
\sqrt{2} Q^i_a \Phi_{ab} Q^j_b J_{ij} + \sqrt{2} m_{ij} Q^i_a Q^j_a
\label{tree}
\eea
where
 $J_{ij}$ is the symplectic metric
$( {0 \atop -1 }{ 1 \atop 0}  ) \otimes {\bf 1}_{N_f \times N_f} $
used to raise and lower $SO(N)$ flavor indices ( $ {\bf 1}_{N_f
\times N_f}$ is the $N_f \times N_f$ identity matrix ) and
$m_{ij}$ is a quark mass matrix $( { 0 \atop 1 }{  1 \atop 0 }  )
\otimes \mbox{diag} ( m_{1}, \cdots, m_{N_f} ) $. Classically, the
global symmetries are the flavor symmetry $Sp(2N_f)$ and $U(1)_R
\times SU(2)_R$ chiral R-symmetry. When $N_f < N-2$, the theory is
asymptotically free and generates dynamically a strong coupling
scale $\La_{N=2}$. The instanton factor is proportional to
$\La_{N=2}^{2N-4-2N_f}$. Then $U(1)_R$ symmetry is anomalous and
broken down to a discrete $Z_{2N-2N_f-4}$ symmetry by instanton.
By taking a tree level superpotential perturbation $\De W$ made
out of the adjoint field in the vector multiplet to the ${\cal
N}=2$ superpotential (\ref{tree}), the ${\cal N}=2$ supersymmetry
can be broken to ${\cal N}=1$ supersymmetry. That is, \bea W =
W_{tree}(\Phi, Q) + \De W, \qquad \De W \equiv \sum_{k=1}^{{\left[
\frac{N}{2} \right]}} \frac{g_{2k}}{2k} \mbox{Tr} \Phi^{2k}. \nonu
\eea Then a microscopic ${\cal N}=1$ $SO(N)$ gauge theory is
obtained from ${\cal N}=2$ $SO(N)$ Yang-Mills theory perturbed by
$\De W$.

By using the perturbed superpotential in addition to the tree
level one, substituting the whole superpotential into the $SO(N)$
matrix model at large $N$ and replacing the gauge theory fields
with matrices, we study the various contributions to the free
energy. Then the partition function can be written as \bea Z =
\frac{1}{vol(SO(N))} \int [d \Phi] [ d Q] \exp
\left[-\frac{1}{g_s}
 W(\Phi) -  \sqrt{2} Q^i_a \Phi_{ab} Q^j_b J_{ij} - \sqrt{2}
m_{ij} Q^i_a Q^j_a \right] \label{Wprime} \eea where for
simplicity we change the notation \bea W(\Phi) =
\sum_{k=1}^{{\left[ \frac{N}{2} \right]+1}} \frac{g_{2k}}{2k}
\mbox{Tr} \Phi^{2k}. \nonu \eea A superpotential $W$ of order
$2(\left[\frac{N}{2}\right]+1)$ breaks the gauge symmetry down to
a direct product of $(\left[\frac{N}{2} \right]+1)$ subgroup. One
can write the derivative of $W$ with respect to the field \bea
W^{\prime}(x) = x^{2N} + \sum_{i=1}^{N} s_{2i} x^{2(N-i)} =
\prod_{i=1}^{N} \left( x^2-e_i^2 \right) \label{Wp}
\eea
where
$e_i$'s are the classical moduli and the symmetric polynomial
$s_{2k}$ in $e_i^2$ is \bea s_{2k} = (-1)^k \sum_{i_1 < \cdots <
i_k} e_{i_1}^2 \cdots e_{i_k}^2. \nonu \eea The description for
the addition of the mass term for the adjoint scalar only was
studied in the context of matrix model \cite{ahn03}. The matrix
description for pure flavors without any adjoint fields to check
the Seiberg duality was observed in \cite{an}. Let us study
$SO(N)$ matrix model by considering even $N$ and odd $N$ case
separately because the Jacobian has different form in each case
and also the spectral curves are different. Now we first analyze
$SO(2N)$ matrix model.

$\bullet$ $SO(2N)$ matrix model

According to the procedure \cite{dv3,mcg,nsw2,ashoketal}
and by integrating over $Q$
in our case, the eigenvalue  basis provides
\bea
Z \sim \int \prod_{a=1}^{N} [d \la] \prod_{a < b}^{N}
\left( \la_a^2 -\la_b^2 \right)^2
\exp \left[
{-\frac{1}{g_s} \sum_{a=1}^{N}  2 W(\la_a)
- \sum_{i=1}^{N_f}  \log \left(\la_a^2 -m_i^2 \right)
 } \right]
\nonu
\eea
where $\pm i \la_a$ are the eigenvalues of $\Phi= ( {0
\atop -1 }{ 1 \atop 0}  ) \otimes \mbox{diag} (\la_1, \cdots,
\la_{N})$ and $m_i$ is a quark mass $( { 0 \atop 1 }{  1 \atop 0 }
) \otimes \mbox{diag} ( m_{1}, \cdots, m_{N_f} )$. Note the factor
2 in the first term of the exponent. The second term comes from
the determinant of $(\Phi+m)$. The new thing in our problem is the
flavor part in the last term. For $SO(2N)$ theory without any
flavors these terms are absent \cite{ashoketal}. After
exponentiating, the effective action for the eigenvalues is given
by
\bea S(\la) = -\sum_{a < b}^N \log \left( \la_a^2 -\la_b^2
\right)^2 + \frac{1}{g_s} \sum_{a=1}^{N} 2 W(\la_a) +
\sum_{i=1}^{N_f}  \log \left(\la_a^2 -m_i^2 \right).
\nonu
\eea
In
this way, the potential $W(\la)$ contains a collection of $N$
variables $\la_1, \la_2, \cdots, \la_{N}$. Remember that due to
the antisymmetric property of $\Phi$ and the trace of it vanishes,
only even terms in the potential $W$ which is a polynomial of
order $(2N+2)$ contribute. The saddle point equations (classical
equations of motion) coming from varying the action with respect
to a single eigenvalue $\la_a$ are
\bea \sum_{b\neq a}^N
\frac{2\la_a}{\la_a^2-\la_b^2} -\frac{1}{g_s} W^{\prime} (\la_a) -
\sum_{i=1}^{N_f} \frac{\la_a}{\la_a^2-m_i^2} =0. \label{saddle}
\eea

To solve this let us introduce the trace of the resolvent
of the matrix
$\Phi$ \cite{fgz,dv3,mcg,nsw2,ashoketal}
\bea
\omega(x) = \frac{1}{N} \mbox{Tr} \frac{1}{\Phi-x} = \frac{1}{N} \sum_{a=1}^
{N} \frac{2x}{\la_a^2-x^2}.
\label{omega}
\eea
Then multiplying
(\ref{saddle}) by $2\la_a/(x^2-\la_a^2)$ and summing over an index $a$,
one gets a loop  equation for $\om(x)$
\bea
\om^2(x)
 +  \frac{2}{S} \om(x)  W^{\prime}(x) + \frac{f(x)}{S^2}=0
\nonu
\eea
where 
the $S$ is defined as $S \equiv g_s N$ being fixed in the large $N$ limit
and 
the polynomial $f(x)$ is given by
\bea
f(x)  \equiv  4 g_s \sum_{a=1}^{N}
 \frac{\la_a  W^{\prime}(\la_a)- x  W^{\prime}(x) }{\la_a^2-x^2}
\nonu \eea which is a polynomial of order $(2N-2)$ with even
powers. Therefore the function $f(x)$ determines the solution of
the matrix integral. Here we take the large $N$ limit and drop
the terms like $\om(x)/x$ and $\om^{\prime}(x)$ which will be
important when we expand it with respect to $1/N$ in order to
derive the Seiberg-Witten differential completely within the
framework of the matrix model.  
The spectral curve reduces to 
\bea
y^2= {W^{\prime}(x)}^2 - f(x), \qquad f(x) = \sum_{n=0}^{N-1}
b_{2n} x^{2n} 
\nonu 
\eea 
where we define \bea y(x) =W^{\prime}(x)+
S \om(x). 
\nonu 
\eea 
This is nothing but a hyperelliptic curve in
$(x,y)$ plane. 
 We have to determine the $N$ unknown
coefficients $b_{2n}$.
As in \cite{cv}, there exists two particular points
denoted by $P$ and $Q$ located at the two pre images of $\infty$
of $x$. The force equation becomes \bea 2 y(\la)= - g_s \frac{\pa
S}{\pa \la}. \nonu \eea The solution for resolvent is given by
\cite{fgz} 
\bea \om(x) =
  \sqrt{{W^{\prime}(x)}^2 - f(x) }- W^{\prime}(x).
\nonu 
\eea 
which is expressed as an $N$ unknown coefficient
function appearing in the polynomial $f(x)$. The resolvent has the
branch cuts among which the eigenvalues of the matrix are
distributed.

 In the large $N$ limit, the distribution of
eigenvalues can be written as \bea \rho(\la)= \frac{1}{N}
\sum_{a=1}^{N} \de(\la -\la_a), \qquad \int \rho(\la) d \la =1
\nonu \eea and the resolvent becomes in this limit \bea \om(x) = 2
\int_{0}^{\infty} \frac{x \rho(\la) d \la }{\la^2-x^2} =
  \int_{0}^{\infty}
\rho(\la) d \la  \left(\frac{1}{\la-x} - \frac{1}{\la+x} \right)=
 \int_{-\infty}^{\infty}
 \frac{\rho(\la) d \la}{\la-x}
\nonu
\eea
which implies that
\bea
\rho(\la)  =
 \frac{1}{2\pi i} \left[ \om(\la +i \ep) - \om(\la-i \ep) \right]
 =  \frac{1}{2\pi i} \left[ y(\la +i \ep) - y(\la-i \ep) \right]. \nonu
\eea
The filling fractions are given by
\bea S_i = \frac{1}{2 \pi
i} \int_{A_i} y d x, \qquad S_0 = \frac{1}{4 \pi i} \int_{A_0} y d
x
\nonu
\eea
where we take the half of the cycle around $A_0$ due
to the orientifold projection. In order to find out the functional
behavior of $f(x)$ within the matrix model, the saddle point
computation of the partition function gives rise to up to $1/g_s$
term as follows:
\bea Z &=& \exp \left[-\frac{2S}{g_s^2} \int d \la
\rho(\la) W(\la) + \frac{S^2}{g_s^2} \int d \la d \la^{\prime}
\rho(\la)
\rho(\la^{\prime}) \log (\la^2 -{\la^{\prime}}^2) \right] \nonu \\
&& +\exp \left[ -\frac{S}{g_s} \sum_{i=1}^{N_f} \int d \la
\rho(\la) \log (\la^2 -m_i^2) \right] \equiv \exp
\left[-\frac{1}{g_s^2} {\cal F}_2 - \frac{1}{g_s} {\cal
F}_1\right]. \label{Z1}
\eea
To get the effective superpotential,
one should know both the variation of ${\cal F}_2$ under a small
change in $S_i$ and ${\cal F}_1$ that can be read off from
(\ref{Z1}).  For the former, we take the following change in
$\rho(\la)$
\bea \rho(\la) \rightarrow \rho(\la) + \frac{\de
S_i}{S} \de (\la -e_i)
\nonu
\eea
where
$e_i$ is an arbitrary point along the $i$-th cut on the hyperelliptic
Riemann surface.
From the explicit form of ${\cal F}_2$ in (\ref{Z1})
one considers
\bea
\de {\cal F}_2 = \de S_i \left( 2W(e_i) - 2S \int d \la \rho(\la)
\log (\la^2-e_i^2)\right)
\nonu
\eea
up to $S_i$ independent terms which are not relevant.
Then the partial derivative of ${\cal F}_2$ with respect to
$S_i$ becomes
\bea
\frac{\pa {\cal F}_2}{\pa S_i } & = & - 2 \int_{e_i}^{P} dx W^{\prime}(x)
-4 S \int d \la \rho(\la)  \int_{e_i}^{P}
\frac{x d x}{\la^2-x^2}
\nonu \\
&=&   - 2  \int_{e_i}^{P}
dx \left( W^{\prime}(x) +  S \om(x)  \right)
=  - 2 \int_{e_i}^{P} y(x)  dx  =
 - \left( \int_{e_i}^{P} +  \int^{e_i}_Q \right) y(x)  dx
\label{F2}
\eea
up to an irrelevant constant of integration terms.
Moreover the ${\cal F}_1$ term can be written as \bea {\cal F}_1 &
= &  2S \sum_{i=1}^{N_f} \int d \la \rho(\la)
 \int_{ m_i}^{P}  \frac{x d x}{ \la^2-x^2}
=   S  \sum_{i=1}^{N_f}
 \int_{m_i}^{P}   \om(x) d x  \nonu \\
& = &   \sum_{i=1}^{N_f}
\int_{ m_i}^{P} y(x) d x = \frac{1}{2} \sum_{i=1}^{N_f}
\int_{ m_i}^{P} y(x) d x +
 \frac{1}{2} \sum_{i=1}^{N_f}
\int_{-m_i}^{P} y(x) d x
\label{F1}
\eea
up to the $S_i$ independent terms.

Combining the two contributions  (\ref{F2}) and (\ref{F1})
one gets the effective superpotential
\bea
W=
 -\frac{1}{2}(2N-2) \int_{Q}^{P} y dx + \frac{1}{2} \sum_{i=1}^{N_f}
\int_{ m_i}^{P} y d x +
 \frac{1}{2} \sum_{i=1}^{N_f}
\int_{-m_i}^{P} y d x+ \cdots
\label{Wimpo}
\eea
Here there are
some remarks in order. 
In the type IIB string theory, the $U(N)$ gauge theory is realized by
the worldvolume of $N$ D5-branes wrapped on ${\bf S}^2$
and in the dual geometry D5-branes are replaced by RR fluxes
generating the effective superpotential and the ${\bf S}^2$ by ${\bf S}^3$. 
In order to deal with the gauge groups $SO(N)$ and  $Sp(N)$, 
one needs to introduce the orientifold plane into the geometry.
This will change the contributions of RR fluxes below.
The physical D5-brane charge of orientifold
plane (O5-plane) is $-1$ and the total $2N_0$ D5-branes wrapping
around the origin should be modified by $(2N_0-2)$. Moreover the
branch cuts in figure 1 in \cite{feng1} are symmetric, due to the
$Z_2$ symmetry, with the one located in the center (We follow the
notations given in \cite{feng1}). This implies that the
contribution from the compact cycle with $a_{-k}$ is exactly the
same as the one with $a_k$. That is, by replacing the D5-branes
with the fluxes, one gets \cite{feng1}
\bea \int_{a_k} h =
\frac{1}{2} N_k=\int_{a_{-k}} h = \frac{1}{2} N_{-k}, \qquad
\int_{a_0} h =\frac{1}{2} (2N_0-2).
\nonu
\eea
By summing over the
all $\al_k$ contour,
\bea \oint_{P} h =\left( 2\sum_{k=1}^N N_k
+2N_0 \right)-2 =(2N-2)-N_f, \qquad \oint_{Q} h = -(2N-2), \qquad
\oint_{\pm m_i} h = -1
\nonu
\eea
where $h$ is an one form on the
Riemann surface. So $h$ should have a pole of order 1 at $P$ and
$Q$ with residues $(2N-2-N_f)$ and $-(2N-2)$ respectively 
\cite{ookouchi}. 
By using
the properties $C_{-k} =-\sum_{j=1}^{k} \be_{-j} +C_0$ and $C_{k}
=\sum_{j=1}^{k} \be_{j} +C_0$ given in \cite{feng1}, one can
divide into two parts: the cycle around $C_0$ and the one around
$\be_j, j=1,2, \cdots, k$. Then we have extra two contributions in
(\ref{Wimpo}) denoted by $\cdots$, Yang-Mills coupling term and
the term with the cycle $\be_j$. But if we take the variation of
$W$, the first contribution will give rise to a trivial cycle and
the second one gives an element of the period lattice. For $U(N)$
gauge theory, the discussion on this matter was considered in
\cite{cv}. Since $S_i$ can be determined by $f(x)$ (and therefore
$b_{2n}$), we have to compute the variation of $W$ with respect to
$b_{2n}$.
\bea \frac{\pa y}{\pa b_{2n}} d x = -\frac{x^{2n}}{2y}
dx, \qquad n=0, 1, \cdots, N-1
\nonu
\eea
which are
basis for the subspace of
holomorphic differentials (one forms) which are odd
under the ${\bf Z}_2 $ transformation $x \rightarrow -x$,  
on the Riemann surface. 
Note that 
the full space of holomorphic differentials has the dimension
$(2N-1)$ due to the genus $(2N-1)$ of the Riemann surface. 

By changing the bases to the homology basis, the extremum
condition of $W$ will give rise to \bea
(2N-2) \int_{p_0}^Q \zeta_k -(2N-2N_f-2)
\int_{p_0}^P \zeta_k -  \sum_{i=1}^{N_f}
\left( \int_{p_0}^{ m_i} +  \int_{p_0}^{- m_i} \right) \zeta_k   =0
\nonu
\eea
modulo the period lattice and $p_0$ is an arbitrary generic
point on the Riemann surface.

There exists a function on the Riemann surface with an $(2N-2)$-th order
pole at $Q$, an $(2N-2N_f-2)$-th order zero at $P$ and
simple
zeros at $x=\pm m_i$ for each
$i=1, 2, \cdots, N_f$, according to Abel's theorem \cite{cv,nsw2}.
The function is simply related to the resolvent divided by $x^2$
(We refer to \cite{ookouchi} with flavors and
\cite{feng1} without flavors for the geometric picture)
\bea
z(x) = \frac{y}{x^2}- \frac{W^{\prime}(x)}{x^2}=
 \sqrt{\frac{{W^{\prime}(x)}^2}{x^4}-\frac{f(x)}{x^4}} -
 \frac{W^{\prime}(x)}{x^2}, \qquad 2N_f < 2N-2.
\nonu \eea This function has an $(2N-2)$-th order pole at $Q$ and
at least fourth order pole at $P$ since $f(x)$ is a polynomial of
at most $(2N-2)$-th order. Therefore $f(x)$ should contain a
factor $(x^2-m_i^2)$ for each $i$ in order for $z(x)$ to have a
simple zero at $x= \pm m_i$ and should be at most $(2N_f+4)$-th
order in order for $z(x)$ to satisfy an $(2N-2N_f-2)$-th order
zero at $P$. Combining these two conditions and putting the
proportional constant as $\La_{N=2}^{4N-2N_f-4}$, the spectral
curve is \bea y^2 = \prod_{i=1}^{N}  (x^2 -e_i^2)^2-
\La_{N=2}^{4N-2N_f-4} x^4 \prod_{i=1}^{N_f} (x^2-m_i^2) \nonu \eea
which is exactly the Seiberg-Witten curve 
\cite{bl,as,hanany,dkp,aot}
in the field theory analysis where we used the relation
(\ref{Wp}). So far we assumed that the number of flavors are not
too large. Next we consider the case of large flavors.

For $2N-2 < 2N_f$, the function $z(x)$ is given by
\bea z(x) = 
\sqrt{\frac{A(x)^2}{x^4}-\frac{g(x)}{x^4}} -
 \frac{A(x)}{x^2}, \qquad  2N-2 < 2N_f < 4N-4
\nonu
\eea
where $A(x)$ is an $2N$-th order polynomial and $g(x)$
is proportional to $x^4 \prod_{i=1}^{N_f} (x^2-m_i^2)$. The
presence of this new function comes from the fact that when the
$2N_f$ is greater than $(2N-2)$, it is not enough to have the
function $f(x)$ only because the pole structure at $P$ needs to
introduce a new function. Now we take the proportional constant as
$\La_{N=2}^{4N-2N_f-4}$. Then the function $z(x)$ vanishes at
$x=\pm m_i$ for each $i=1, 2, \cdots, N_f$ and has an $(2N-2)$-th
order pole at $Q$ and $(2N_f-2N+2)$-th order pole at $P$. The
expression inside of square root in $z(x)$ should be proportional
to $y(x)^2/x^4$
\bea \frac{A(x)^2}{x^4} - \La_{N=2}^{4N-2N_f-4}
\prod_{i=1}^{N_f} (x^2-m_i^2) =
\frac{{W^{\prime}(x)}^2}{x^4}-\frac{f(x)}{x^4}
\nonu
\eea
where
$f(x)$ is a polynomial of order at most $(2N-2)$. The solution for
this up to ${\cal O}(\La_{N=2}^{4N-2N_f-4})$ is given by
\bea A(x)
& = & \prod_{i=1}^{N} (x^2 -e_i^2) + \La_{N=2}^{4N-2N_f-4}
P(x), \nonu \\
f(x) & = & \La_{N=2}^{4N-2N_f-4} \left( x^4 \prod_{i=1}^{N_f}
(x^2-m_i^2)- 2 P(x) \prod_{i=1}^{N}  (x^2 -e_i^2)\right) \nonu
\eea
where $ P(x)$ is defined as a polynomial of degree
$(2N_f-4N+4)$ in $x$ and $m_i$. Therefore the spectral curve and
the function are given by \bea
y^2  & = &   \prod_{i=1}^{N}  (x^2 -e_i^2)^2 -f(x) \nonu \\
& = &
 \prod_{i=1}^{N}  \left[ (x^2 -e_i^2) + \La_{N=2}^{4N-2N_f-4}
P(x)  \right]^2- \La_{N=2}^{4N-2N_f-4} x^4
\prod_{i=1}^{N_f} (x^2-m_i^2), \nonu \\
z(x) & = &   \frac{y}{x^2} -\frac{A(x)}{x^2}. \nonu \eea
Therefore in both regions of the number of flavors the spectral
curve coming from the matrix model calculations coincides with
precisely the known Seiberg-Witten curve.

 $\bullet$ $SO(2N+1)$ matrix model

Let us describe the odd case. Since the presentation looks similar to
the one in previous discussion, we will present the main difference only.
By integrating over $Q$, the eigenvalue  basis
provides \cite{dv3,mcg,nsw2,ashoketal}
\bea
Z \sim \int \prod_{a=1}^{N} [d \la] \prod_{a < b}^{N}
\left( \la_a^2 -\la_b^2 \right)^2 \prod_{a=1}^{N} \la_a^2
\exp \left[
{-\frac{1}{g_s} \sum_{a=1}^{N}  2 W(\la_a)
- \sum_{i=1}^{N_f}  \log \left(\la_a^2 -m_i^2 \right)
 } \right]
\nonu
\eea
where $\pm i \la_a$ are the eigenvalues of $\Phi= ( {0
\atop -1 }{ 1 \atop 0}  ) \otimes \mbox{diag} (\la_1, \cdots,
\la_{N},0)$ and $m_i$ is a quark mass $( { 0 \atop 1 }{  1 \atop 0
}  ) \otimes \mbox{diag} ( m_{1}, \cdots, m_{N_f} )$. Note that
there exists an extra factor $ \prod_{a=1}^{N} \la_a^2 $ in this
case \cite{ashoketal}. Then the effective action for the
eigenvalues is given by
\bea S(\la) = -\sum_{a < b}^N \log \left(
\la_a^2 -\la_b^2 \right)^2 - \sum_{a=1}^N \log \la_a^2 +
\frac{1}{g_s} \sum_{a=1}^{N}  2 W(\la_a) + \sum_{i=1}^{N_f}  \log
\left(\la_a^2 -m_i^2 \right).
\nonu
\eea
By varying the action
with respect to an eigenvalue one gets
\bea \sum_{b\neq a}
\frac{2\la_a}{\la_a^2-\la_b^2} -\frac{1}{\la_a} -\frac{1}{g_s}
W^{\prime} (\la_a) -  \sum_{i=1}^{N_f} \frac{\la_a}{\la_a^2-m_i^2}
=0.
\nonu
\eea
In the large $N$ limit, the second extra term
comparing with the previous case $SO(2N)$ matrix model does not
contribute. Therefore all the arguments from (\ref{omega}) to
(\ref{F1}) in even $SO(2N)$ gauge theory are valid for $SO(2N+1)$
matrix model. By an appropriate counting the physical D5-brane
charge of the orientifold which is equal to $-\frac{1}{2}$, one
gets
\bea \int_{a_k} h = \frac{1}{2} N_k=\int_{a_{-k}} h =
\frac{1}{2} N_{-k}, \qquad \int_{a_0} h =\frac{1}{2} (2N_0-1).
\nonu
\eea
By summing over the all $\al_k$ contour,
\bea \oint_{P}
h =\left( 2\sum_{k=1}^N N_k +2N_0 \right)-1 =(2N-1)-N_f, \qquad
\oint_{Q} h = -(2N-1), \qquad \oint_{\pm m_i} h = -1
\nonu
\eea
where $h$ is an one form on the Riemann surface.

The extremum condition of $W$ (\ref{Wimpo}) where the coefficient
$(2N-2)$ is replaced with $(2N-1)$ will give rise to \bea (2N-1)
\int_{p_0}^Q \zeta_k -(2N-2N_f-1) \int_{p_0}^P \zeta_k -
\sum_{i=1}^{N_f} \left( \int_{p_0}^{ m_i} +  \int_{p_0}^{- m_i}
\right) \zeta_k =0 \nonu \eea modulo the period lattice. Then the
function is simply related to the resolvent divided by $x$ \bea
z(x) = \frac{y}{x}- \frac{W^{\prime}(x)}{x}= 
\sqrt{\frac{{W^{\prime}(x)}^2}{x^2}-\frac{f(x)}{x^2}} -
 \frac{W^{\prime}(x)}{x}, \qquad
2N_f < 2N-1.
\label{res}
\eea
This function has an $(2N-1)$-th
order pole at $Q$ and at least third order pole at $P$ since
$f(x)$ is a polynomial of at most $(2N-2)$-th order. Therefore
$f(x)$ should contain a factor $(x^2-m_i^2)$ for each $i$ in order
for $z(x)$ to have a simple zero at $x= \pm m_i$ and should be at
most $(2N_f+2)$-th order in order for $z(x)$ to satisfy an
$(2N-2N_f-1)$-th order zero at $P$. Combining these two conditions
and putting the proportional constant as $\La_{N=2}^{4N-2N_f-2}$,
the spectral curve is
\bea y^2 = \prod_{i=1}^{N}  (x^2 -e_i^2)^2-
\La_{N=2}^{4N-2N_f-2} x^2 \prod_{i=1}^{N_f} (x^2-m_i^2)
\nonu
\eea
which is known as the Seiberg-Witten curve
\cite{ds,as,hanany,dkp,aot} in the perturbative calculation in gauge
theory side.

For $2N-1 < 2N_f$, the function $z(x)$ is given by
\bea z(x) = 
\sqrt{\frac{A(x)^2}{x^2}-\frac{g(x)}{x^2}} -
 \frac{A(x)}{x}, \qquad  2N-1 < 2N_f < 4N-2
\label{res1}
\eea
where $A(x)$ is an $2N$-th order polynomial and
$g(x)$ is proportional to $x^2 \prod_{i=1}^{N_f} (x^2-m_i^2)$. Now
we take the proportional constant as $\La_{N=2}^{4N-2N_f-2}$. Then
the function $z(x)$ vanishes at $x=\pm m_i$ for each $i=1, 2,
\cdots, N_f$ and has an $(2N-1)$-th order pole at $Q$ and
$(2N_f-2N+1)$-th order pole at $P$. The expression inside of
square root in $z(x)$ should be proportional to $y(x)^2/x^2$
\bea
\frac{A(x)^2}{x^2} - \La_{N=2}^{4N-2N_f-2} \prod_{i=1}^{N_f}
(x^2-m_i^2) = \frac{{W^{\prime}(x)}^2}{x^2}-\frac{f(x)}{x^2} \nonu
\eea
where $f(x)$ is a polynomial of order at most $(2N-2)$. It is
straightforward to see the coincidence of the Seiberg-Witten
curve. From the properties of $z(x)$ in (\ref{res}) and
(\ref{res1}), one can easily see that \bea h(x) d x= \frac{d z}{z}
\nonu \eea is a meromorphic differential with simple poles at $P,
Q$ and $x=\pm m_i$ with residues $(2N-2N_f-2), -(2N-2)$ and 1
respectively for $SO(2N)$ gauge theory, for example.

\section{$Sp(N)$ matrix model}
\setcounter{equation}{0}

In this section, we continue to study the matrix model for the
symplectic group $Sp(N)$. Let us consider  an ${\cal N}=2$
supersymmetric $Sp(N)$ gauge theory with $N_f$ flavors of quarks
$Q^i_a(i=1, 2, \cdots, 2N_f, a=1, 2, \cdots, 2N)$ in the
fundamental representation. The tree level superpotential of the
theory is obtained from \cite{ty,kty,aps,as,hms,ahn98}
\bea
W_{tree}(\Phi, Q) = \sqrt{2} Q^i_a \Phi^a_b Q^i_c J^{bc} +
\sqrt{2} m_{ij} Q^i_a Q^j_b J^{ab} \label{treesp}
\eea
where
 $J_{ab}$ is the symplectic metric
$( {0 \atop 1 }{ -1 \atop 0}  ) \otimes {\bf 1}_{N \times N} $ and
$m_{ij}$ is a quark mass matrix $( { 0 \atop -1 }{  1 \atop 0 }  )
\otimes \mbox{diag} ( m_{1}, \cdots, m_{N_f} ) $. Classically, the
global symmetries are the flavor symmetry $O(2N_f)$ and $U(1)_R
\times SU(2)_R$ chiral R-symmetry. When $N_f < 2N+2$, the theory
is asymptotically free  and generates dynamically a strong
coupling scale $\La_{N=2}$. The instanton factor is proportional
to $\La_{N=2}^{2N+2-N_f}$. Then $U(1)_R$ symmetry is anomalous and
broken down to a discrete $Z_{2N-N_f+2}$ symmetry by instanton.

According to the procedure \cite{dv3,mcg,nsw2,ashoketal}
and by integrating over $Q$
in our case, the eigenvalue  basis provides
\bea
Z \sim \int \prod_{a=1}^{N} [d \la] \prod_{a < b}
\left( \la_a^2 -\la_b^2 \right)^2
\prod_{a=1}^{N} \la_a^2
\exp \left[
{-\frac{1}{g_s} \sum_{a=1}^{N}  2 W(\la_a)
+ \sum_{i=1}^{N_f}  \log \left(\la_a^2 -m_i^2 \right)
 } \right]
\nonu
\eea
where $\pm i \la_a$ are the eigenvalues of $\Phi=
( {1 \atop 0 }{ 0 \atop -1}  ) \otimes \mbox{diag}
(\la_1, \cdots, \la_{N })$ and
$m_i$ is a quark mass
$( { 0 \atop -1 }{  1 \atop 0 }  )
\otimes \mbox{diag} ( m_{1}, \cdots, m_{N_f} )$.
%
%
%
By an appropriate counting the physical D5-brane charge of the
orientifold which is equal to 1, one gets
\bea \int_{a_k} h =
\frac{1}{2} N_k=\int_{a_{-k}} h = \frac{1}{2} N_{-k}, \qquad
\int_{a_0} h =\frac{1}{2} (2N_0+2).
\nonu
\eea
By summing over the
all $\al_k$ contour,
\bea \oint_{P} h =\left( 2\sum_{k=1}^N N_k
+2N_0 \right)+2 =(2N+2)-N_f, \qquad \oint_{Q} h = -(2N+2), \qquad
\oint_{\pm m_i} h = -1
\nonu
\eea
where $h$ is an one form on the
Riemann surface.

The extremum condition of $W$ (\ref{Wimpo}) where the coefficient
$(2N-2)$ is replaced with $(2N+2)$ will give rise to \bea
(2N+2) \int_{p_0}^Q \zeta_k -(2N-2N_f+2) \int_{p_0}^P \zeta_k -
\sum_{i=1}^{N_f} \left( \int_{p_0}^{ m_i} +  \int_{p_0}^{- m_i}
\right) \zeta_k   =0.
 \nonu
 \eea
 There exists a function on the
Riemann surface with an $(2N+2)$-th order pole at $Q$, an
$(2N-2N_f+2)$-th order zero at $P$ and simple zeros at $x=\pm m_i$
for each $i=1, 2, \cdots, N_f$. The function is simply related to
the resolvent  \cite{ookouchi,feng1}
\bea z(x) =  \sqrt{ x^4
{W^{\prime}(x)}^2-  f(x)} -
 x^2  W^{\prime}(x).
\nonu
\eea
This function has an $(2N+2)$-th order pole at $Q$ and
at least zero-th order pole at $P$ since $f(x)$ is a polynomial of
at most $(2N-2)$-th order. Therefore $f(x)$ should contain a
factor $(x^2-m_i^2)$ for each $i$ in order for $z(x)$ to have a
simple zero at $x= \pm m_i$ and should be at most $2N_f$-th order
in order for $z(x)$ to satisfy an $(2N-2N_f+2)$-th order zero at
$P$. Combining these two conditions and putting the proportional
constant as $\La_{N=2}^{4N-2N_f+4}$, the spectral curve is
\bea y^2 = \left( x^2 \prod_{i=1}^{N}  ( x^2 -e_i^2) \right)^2 -
\La_{N=2}^{4N-2N_f+4} \prod_{i=1}^{N_f} (x^2-m_i^2).
\nonu
\eea
In
the gauge theory side, for the Seiberg-Witten curve
\cite{as,dkp,ahn98} for $Sp(N)$ case at least one hypermultiplet
of exactly zero mass (for example, $m_{N_f}=0$), the above matrix
model result is exactly the same  the one in \cite{as,dkp,ahn98}
because the extra piece which is peculiar to $Sp(N)$ case, the
product of quark mass term vanishes. For nonzero quark
mass, one can consider the following resolvent
$z(x) =  \sqrt{ \left( x^2
{W^{\prime}(x)} + g \right)^2-  f(x)} -
 \left( x^2  W^{\prime}(x) + g \right)$ by including the constant term 
($x$ independent term) $g$ inside the square.  
By identifying this $g$ with $\La_{N=2}^{2N+2-N_f} \prod_{i=1}^{N_f} m_i$,
one obtains the Seiberg-Witten curve.


\vspace{1cm}
\centerline{\bf Acknowledgments}

This research of CA was supported by
Korea Research Foundation Grant(KRF-2002-015-CS0006).
This research of SN was supported by
Korea Research Foundation Grant KRF-2001-041-D00049.
We thank Korea Institute for Advanced Study (KIAS)
where part of this work was undertaken.
CA thanks Y. Ookouchi for the correspondence on his paper.

\end{document}